\documentclass{pkas}

\def\year{2014} 
\def\month{May} 
\def\volume{00} 
\def\issue{0} 
\def\beginpage{1} 
\def\endpage{99} 
\def\received{September 30, 2014} 
\def\accepted{Octover 31, 2014} 
\date{Received \received ; accepted \accepted}

\title{
Hyper Suprime-Camera survey of the AKARI NEP Wide field
}

\author[1]{Tomotsugu Goto
}
\author[2]{Yoshiki Toba}
\author[3]{Yousuke Utsumi}
\author[4]{Nagisa Oi}
\author[4]{Toshinobu Takagi}
\author[5]{Matt Malkan}
\author[6]{Youichi Ohayma}
\author[4]{Kazumi Murata}
\author[7]{Paul Price}
\author[8]{Marios Karouzos}
\author[4]{Hideo Matsuhara}
\author[4]{Takao Nakagawa}
\author[4]{Takehiko Wada}
\author[9]{Steve Serjeant}
\author[10]{Denis Burgarella}
\author[10]{Veronique Buat}
\author[11]{Masahiro Takada}
\author[12]{Satoshi Miyazaki}
\author[13]{Masamune Oguri}
\author[14]{Takamitsu Miyaji}
\author[]{Shinki Oyabu}
\author[9]{Glenn White}
\author[15]{Tsutomu Takeuchi}
\author[]{Hanae Inami}
\author[9]{Chris Perason}
\author[15,16]{Katarzyna Malek}
\author[15]{Lucia Marchetti}
\author[8]{Lee HyungMoK }
\author[8]{Myung Im}
\author[8]{Seong Jin Kim}
\author[1]{Ekaterina Koptelova}
\author[1]{Dani Chao}
\author[1]{Yi-Han Wu}
\author[4]{AKARI NEP Survey team}
\author[4]{AKARI All Sky Survey Team}

\affil[1]{Institute of Astronomy, National Tsing Hua University, No. 101, Section 2, Kuang-Fu Road, Hsinchu, Taiwan 30013, R.O.C; \email{tomo@phys.nthu.edu.tw}}
\affil[2]{Research Center for Space and Cosmic Evolution, Ehime University, Bunkyo-cho, Matsuyama 790-8577, Japan}
\affil[3]{
Hiroshima University, 1-3-1 Kagamiyama, Higashi-Hiroshima, Hiroshima 739-8526, Japan}
\affil[4]{Institute of Space and Astronautical Science, JAXA, Sagamihara, Kanagawa 252-5210, Japan}
\affil[5]{Department of Physics and Astronomy, UCLA, Los Angeles, CA, 90095-1547, USA\\}
\affil[6]{Institute of Astronomy and Astrophysics Academia Sinica, Taipei 10617, Taiwan, Republic of China}
\affil[7]{Princeton University, Department of Astrophysical Sciences, Princeton, NJ 08544, USA}
\affil[8]{Department of Physics \& Astronomy, FPRD, Seoul National University, Shillim-Dong, Kwanak-Gu, Seoul 151-742, Korea}
\affil[9]{Department of Physical Sciences, The Open University, Milton Keynes, MK7 6AA, UK}
\affil[10]{Aix-Marseille  Universit\'e,  CNRS, LAM (Laboratoire d'Astrophysique de Marseille) UMR7326,  13388, Marseille, France}
\affil[11]{Kavli Institute for the Physics and Mathematics of the Universe, The University of Tokyo, Chiba 277-8582, Japan}
\affil[12]{National Astronomical Observatory of Japan, Mitaka, Tokyo 181-8588, Japan}
\affil[13]{Department of Physics, University of Tokyo, 7-3-1 Hongo, Bunkyo-ku, Tokyo 113-0033, Japan}
\affil[14]{Instituto de Astronomía, Universidad Nacional Autónoma de México, Ensenada, Baja California, Mexico}
\affil[15]{Division of Particle and Astrophysical Science, Nagoya University, Furo-cho, Chikusa-ku, Nagoya 464-8602, Japan}
\affil[16]{National Centre for Nuclear Research, ul. Hoża 69, 00-681 Warszawa, Poland}

\def\corrauthor{
T. Goto
}

\def\runningauthor{
T. Goto
}

\def\runningtitle{
Hyper Suprime-Cam survey of the AKARI NEP wide field
}

\def\keywords{
Infrared galaxies, AKARI, Hyper Suprime-Cam
}

\def\abstracttext{
The extragalactic background suggests half the energy generated by stars reprocessed into the infrared (IR) by dust. At z$\sim$1.3, 90\% of star formation is obscured by dust. To fully understand the cosmic star formation history, it is critical to investigate infrared emission. AKARI has made deep mid-IR observation using its continuous 9-band filters in the NEP field (5.4 deg$^2$), using $\sim$10\% of the entire pointed observations available throughout its lifetime. However, there remain 11,000 AKARI's infrared sources undetected with the previous CFHT/Megacam imaging ($r\sim$25.9ABmag). Redshift and IR luminosity of these sources are unknown. These sources may contribute significantly to the cosmic star-formation rate density (CSFRD). For example, if they all lie at 1$<z<$2, the CSFRD will be twice as high at the epoch. 
 We  are carrying out deep imaging of the NEP field in 5 broad bands ($g,r,i,z,$ and $y$) using Hyper Suprime-Camera (HSC), which has 1.5 deg field of view in diameter on Subaru 8m telescope.
This will provide photometric redshift information, and thereby IR luminosity for the previously-undetected 11,000 faint AKARI IR sources. Combined with AKARI's mid-IR AGN/SF diagnosis, and accurate mid-IR luminosity measurement, this will allow a complete census of cosmic star-formation/AGN accretion history obscured by dust.
}

\begin{document}
\pkashead 


\section{Undetected AKARI sources:\label{sec:intro}}

Nature hides; the more intense star-formation, the more obscured. 
The extragalactic background suggests at least half the energy generated by stars has been 
reprocessed into the infrared (IR) by dust \citep{1999A&A...344..322L}.
At z$\sim$1.3, 90\% of star formation is obscured by dust \citep{2005ApJ...632..169L, Goto2010NEP}.
Therefore, a full understanding of the cosmic star formation
history inevitably needs an IR perspective.

The AKARI space telescope has performed a deep mid-infrared imaging survey in the NEP region \citep{2009PASJ...61..375L}.
We are studying the multi-band data of these mid-IR galaxies as shown in Table 1 \citep{Takagi_PAH,Goto2010NEP}. However, because  very dusty objects cannot be detected in the relatively shallow CFHT imaging data \citep[$r<$25.9ABmag;][]{2007ApJS..172..583H,2010ApJS..190..166J,Oi2014}, there remain 11,000 AKARI sources undetected in the optical. 
As a result, we lack an understanding of the redshift and IR luminosity of these sources, i.e., they have been excluded from the past cosmic star formation history (CSFH) analysis. These sources can change our view of CSFH---if they all lie at 1$<z<$2, they will $double$ the cosmic star formation density at that epoch. 

\section{Uniqueness of AKARI mid-IR data:\label{sec:uniq}}
The AKARI NEP is one of the best fields for this investigation,
due to the availability of continuous 9-band mid-IR filters. Spitzer lacks filters between 8 and 24$\mu$m  (the critical wide gap between IRAC and MIPS, excluding the tiny IRS peak up array at 16 $\mu$m). Similarly WISE also has a wide gap between 4 and 12$\mu$m filters. Therefore, no other telescope can provide continuous 9-band photometry in mid-IR wavelength (2-24$\mu$m) in foreseeable future. 
AKARI's continuous  9-band photometry works as a low-resolution spectrum, 
which is critically important for the following key aspects:

\begin{itemize} \item  
 Two physical processes produce the mid-IR emission: hot dust around an AGN, and PAH emission from star-formation. Quantitatively separating these is of fundamental importance. The continuous 9 filters of AKARI have made this possible through precise SED fitting \citep[See examples in ][]{Takagi_PAH, 2014ApJ...784..137K}. Importantly, this is independent of extinction.
 \item 

 Accurately measuring the mid-IR emission line strength (PAH 7.7 $\mu$m) and continuum luminosity. 
Using the 9-band photometry as a low-resolution spectra, 
Oyama et al. (2014, submitted) demonstrated that photometric PAH 7.7 $\mu$m line measurements agree well with spectroscopic ones. 

\end{itemize}
 Neither of these is possible if there is a large gap between mid-IR filters. Therefore, AKARI NEP is the only field, where the two different astrophysical 
power sources can be separated for thousands of IR galaxies, including those with heavy extinction.

\begin{figure}[tbp] 
    \centering
    \includegraphics[angle=0,width=3.1in]{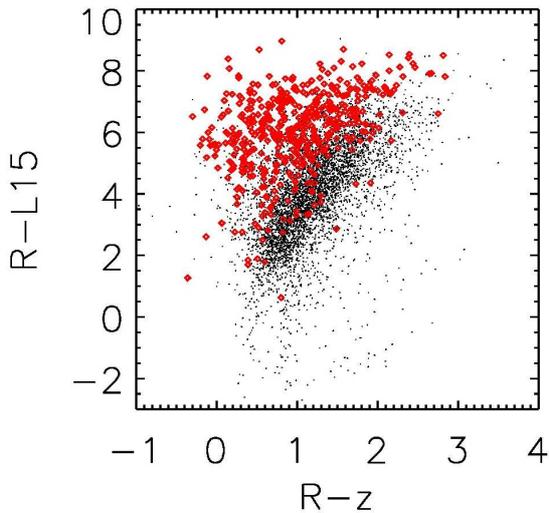} 
    \caption{$R-L15$ color is plotted against $R-z$ color. Black dots are AKARI IR galaxies detected with CFHT/Megacam. Red dots are those not detected with CFHT, but with Suprime-Cam in the central NEP deep field.}
    \label{fig:filterdesign}
 \end{figure}

\begin{figure*}[tbp] 
    \centering
    \includegraphics[angle=0,width=4.2in]{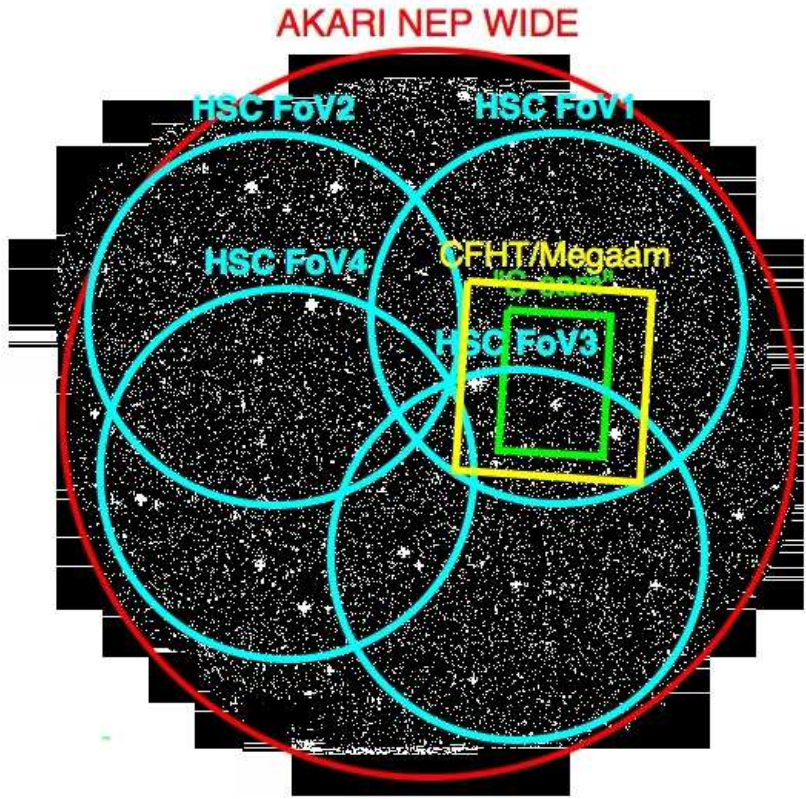} 
    \caption{HSC $r$-band data shown on the background covers the entire NEP wide field (5.4 deg$^2$). 
 Overlaid cyan circles are planned HSC pointings for the remaining filters ($g,r,i,z,$ and $y$). 
Central Suprime-Cam and Megacam fields, where we already have deep optical imaging, are shown in green and yellow.}
    \label{fig:FoV}
 \end{figure*}

The AKARI NEP also has been thoroughly observed in every other available waveband (Table 1), making it one of the premier large deep fields on the sky.
Its large area overcomes the serious problem of cosmic variance, which hampered previous IR CSFH studies. 
In particular, Spitzer's CDFS field was only $~0.25$ deg$^2$  \citep{2005ApJ...632..169L}, and measured an IR luminosity density nearly a factor of 10 different from other Spitzer fields \cite[e.g.,][]{2006MNRAS.370.1159B}. For the same reason, the single Suprime-Cam pointing in the center of the NEP deep field ($0.25$ deg$^2$) is not wide enough.
A large volume coverage also allows us to study environmental effects on galaxy evolution  \citep{2008MNRAS.391.1758K,cluster_LF}.
AKARI was a survey telescope, which observed 5.4 deg$^2$ in NEP using $\sim10$\% of the entire pointed observations available throughout the lifetime of the mission,  providing uniquely precious space-based IR data spanning a large enough area to overcome cosmic variance.

\begin{table*}[tb]
\begin{center}
\caption{Summary of AKARI NEP survey data}\label{tab:multi}
\begin{tabular}{lccl}
\hline 
Observatory  & Band              &   Sensitivity/Number of objects/exposure time          & Area (deg$^2$)\\
\hline  
AKARI/IRC    & {\bf 2.5-24$\mu$m}     &  $L15$=18.5AB  & {\bf 5.4}\\
Subaru/S-Cam  & $BVRi'z'$ & $R$=27.4AB   & 0.25\\
Subaru/FOCAS & optical spect.,   & 57 sources in NEP  &   $R\sim 24$ AB \\
MMT6m & optical spec.       & $\sim$1800 obj  &      5.4            \\
KPNO-2.1m & $J, H$  & 21.6,21.3 & 5.4\\
Maidanak 1.5m  & $B,R,I$  & $R$=23.1 & 3.4\\
KPNO2m/FLAMINGOS  & $J,H$  & $J$=21.6, $H$=21.3 & 5.4\\
WIRCAM & Y,J, K & 24AB  & 1\\
MegaCam & $ugriz$                   &  $ r\leq25.9$AB    & 1   \\
GALEX        & NUV, FUV              & NUV=26     & 1.5       \\
WSRT         & 20cm           &  $\sim$100~$\mu$Jy & 0.25\\
VLA-archive  & 10cm                  & 200~$\mu$Jy       & 5.4 \\
GMRT         & 610MHz                   & 60-80~$\mu$Jy   & 0.25        \\
Keck/Deimos         & optical spec.     & $\sim$1000 obj    & 0.25      \\
Subaru/FMOS         & near-IR spec.       & $\sim$700 obj   & 0.25       \\
   Herschel  & 100,160 $\mu$m  & 5-10 mJy & 0.5\\
        Herschel  & 250-500 $\mu$m & $\sim$10 mJy & 7.1\\
Chandra      & X-ray  &    30-80ks & 0.25\\
SCUBA2      & submm  &    1mJy  & 0.25\\
{\bf Subaru/HSC}      & $r$  &    $r$=27.2(Fig.\ref{fig:FoV})  & {\bf 5.4}\\
\hline 
\end{tabular}
\end{center}
\end{table*}

\section{Subaru Hyper Suprime-Cam survey:\label{sec:hsc}}
To rectify the situation, we are carrying out an optical survey of the AKARI NEP wide field using Subaru's new Hyper Suprime-Cam \cite[HSC;][]{2012SPIE.8446E..0ZM}.
  The HSC has a field-of-view (FoV) of 1.5 deg in diameter, covered with 104 red-sensitive CCDs.
 It has the largest FoV among optical cameras on an 8m telescope, and  can cover the AKARI NEP wide field (5.4 deg$^2$) with only 4 FoV (Fig.\ref{fig:FoV}).

Our immediate science aim of the optical survey is to detect all AKARI sources in the optical, with photometry accurate enough for reliable photometric redshifts.
This allows us to determine the optical and
IR luminosities (corresponding to direct and dust-obscured emission)
from young stars and accreting massive black holes for a large sample representative of the cosmic history of the Universe.

Our main targets--that cannot be detected or measured accurately with smaller telescopes (e.g. CFHT)--
are very dusty galaxies with $r-L15\sim$8. AKARI's sensitivity in NEP wide area is $L15$=18.5ABmag \citep{2009PASJ...61..375L}. Therefore, optical data need to reach $r\sim$26.5mag.
In Fig.\ref{fig:filterdesign}, black dots are bright IR sources detected with CFHT/Megacam. The red points show colors of AKARI IR sources that can only be detected with 8m depth in the central NEP deep region \citep{2008PASJ...60S.531G}. There exist 10 times more of these optically-undetected IR sources in the entire AKARI NEP area \citep{2012A&A...548A..29K}.
 We derived required deer observations in other bands so that SEDs of Arp220, Mrk231, and M82 (with $L15$=18.5) can be detected at z=0.5-1. They are  27.5, 25.5, 24.7, and 24.4 mag in $g,i,z,$ and $y$, respectively. 
If we used a 4m telescope, this would take $\sim~$120 hours in total ($\times$4 more exposure time, $\times$1.7 difference in FoV). Therefore, the Subaru/HSC is the only feasible instrument for this observation.
Since our targets are heavily obscured, depths in bluer bands are important. 

 We have abundant spectroscopic data to calibrate photometric redshifts ($\sim$1000 from Keck/Deimos, and $\sim$1000 near-IR spectra from Subaru/FMOS, $\sim$1800 from MMT/Hetrospec \citep{2013ApJS..207...37S} as summarized in Table \ref{tab:multi}. 
 Our experience in the central Suprime-cam field predicts photo-z accuracy of $\frac{\Delta z}{1+z}$
$\sim$0.036 is achievable from 5-broad band photometry \citep{Oi2014}.

\section{Legacy value of the HSC data:\label{sec:legacy}}

The HSC optical data in the NEP region will have crucial added legacy value. 
ESA's 1.2m Euclid space telescope (planned launch in 2019), for which several of us are members (T.G., S.S., C.P.), will survey  
large deep fields to near-IR 26ABmag, with simultaneous near-IR spectroscopy ($R=250$). These will be exquisite data sets for galaxy/QSO studies to z=8 and beyond, and of Type Ia supernovae to z=1.5. However, orbital constraints strongly restrict the field choices to within 5 deg of the ecliptic poles. One disadvantage of Euclid is it has only one broad filter in the optical. Therefore, HSC's deep 5 broad-band optical photometry is crucial to compliment the deep near-IR data from Euclid. Without HSC data, Euclid's deep near-IR data will not even have decent photometric redshifts.

 Due to similar orbital constraints,  the eROSITA space telescope will also have an exceptionally deep X-ray field around the NEP (scheduled launch in 2015). 
The NEP is also a promising deep field candidate for SPICA,
 the next generation 3.2m IR space telescope with far-infrared wide-field IFU capability. NASA's JWST will also have excellent NEP visibility. For long-period variable and slow moving objects, taking data at early epoch will only increase the value. 
Finally, the NEP is also part of the LOFAR tiered 30,60,120,200MHz extragalactic survey, with a 120MHz RMS$\leq$25$\mu$Jy. 
The NEP field may not have as rich ancillary data as COSMOS or CDFS at the moment, but it will in the next 10 years, because of its visibility from Space.
 Therefore, HSC data are not only important for AKARI, but will continue to have a legacy value for at least the next decade.

\section{2014 HSC observing run:\label{sec:obs}}
Our proposal to the Subaru telescope was accepted with a high score (8.0 out of 10). 
We were awarded two nights of Subaru time on June 29 and 30, 2014. However due to instrumental problems such as  a crash of the instrument control PC, we were only able to observe on June 30, without the filter exchanger, i.e., we were limited to the $r$-band observation. 

 Nevertheless, we carried out the observation in $r$-band. 
 Due to a problem with a dome air control during the daytime, the seeing was 1.5-2.0'', while out side seeing was 0.4-0.6''. Because the temperature were higher inside the dome (6C vs 4C outside), we tried to open the ventilation windows, knowing this increases the turbulence within the dome. This helped to improve the seeing from 2.0'' to 1.5''. However, as soon as we closed the ventilation windows, the seeing went back to 2.0''. We decided to keep the ventilation windows open for the rest of the night.
 
 We observed in 7 FoVs with sets of 5 point dithering pattern. 
 Because the largest spatial gaps of the HSC CCDs are $\sim$53'', we shifted $\sim$120'' at each dither. This is to ensure to fill CCD gaps and to have enough stars observed in two separate CCDs for calibration purpose. Our dither pattern were designed so that $\Delta$RA, and $\Delta$DEC distributions become random, with no repeated points.

\section{Data reduction:\label{sec:reduction}}

 Due to its large size, the reduction of HSC data is a challenge. With the courtesy of the HSC team and NAOJ, we used the HSC pipeline developed by the team. 
 Over 200 GB of data taken in one night, it took 7 days of cputime to reduce and mosaic data with a workstation with dual CPUs (E5-2620-v2 6core, 2.1 Ghz, 24 threads with Hyper Threading Technology in total) and 64GB of memory. Depth varies across the field due to the seeing variation during the night. However, in deepest portions we reached $\sim$27AB mag.

\section{Summary:\label{sec:summary}}
We are carrying out deep HSC $g,r,i,z,$ and $y$ imaging of the AKARI NEP wide field (5.4 deg$^2$). The  HSC observations will provide photometric redshifts and thereby IR luminosities for 11,000 faint AKARI IR sources, previously undetected with CFHT. Combined with AKARI's mid-IR AGN/SF diagnosis, and accurate luminosity measurement from the unique 9 mid-IR photometry, this will provide a complete census of the obscured cosmic star-formation/AGN accretion history at high-z, to be compared with the state-of-the-art low-z work also from AKARI \citep{Goto2011IRAS,Goto2011SDSS,Ece2014}.

\vskip .2cm 
\noindent{\bf Acknowledgments:}
We acknowledge the support by the Ministry of Science and Technology of Taiwan through grant
NSC 103-2112-M-007-002-MY3.
This work was supported by JSPS KAKENHI Grant Number 26800103

\bibliography{201404_goto} 

\begin{thebibliography}{19}
\expandafter\ifx\csname natexlab\endcsname\relax\def\natexlab#1{#1}\fi

\bibitem[{Babbedge} et~al.(2006){Babbedge}, {Rowan-Robinson}, {Vaccari}
  et~al.]{2006MNRAS.370.1159B}
{Babbedge} T.~S.~R., {Rowan-Robinson} M., {Vaccari} M., et~al., 2006, \mnras,
  370, 1159

\bibitem[{Goto} et~al.(2011{\natexlab{a}}){Goto}, {Arnouts}
  et~al.]{Goto2011SDSS}
{Goto} T., {Arnouts} S.~{Malkan} M., et~al., 2011{\natexlab{a}}, \mnras, 414,
  1903

\bibitem[{Goto} et~al.(2011{\natexlab{b}}){Goto}, {Arnouts}, {Inami}
  et~al.]{Goto2011IRAS}
{Goto} T., {Arnouts} S., {Inami} H., et~al., 2011{\natexlab{b}}, \mnras, 410,
  573

\bibitem[{Goto} et~al.(2008){Goto}, {Hanami}, {Im} et~al.]{2008PASJ...60S.531G}
{Goto} T., {Hanami} H., {Im} M., et~al., 2008, \pasj, 60, 531

\bibitem[{Goto} et~al.(2010{\natexlab{a}}){Goto}, {Koyama}, {Wada}
  et~al.]{cluster_LF}
{Goto} T., {Koyama} Y., {Wada} T., et~al., 2010{\natexlab{a}}, \aap, 514, A7

\bibitem[{Goto} et~al.(2010{\natexlab{b}}){Goto}, {Takagi}, {Matsuhara}
  et~al.]{Goto2010NEP}
{Goto} T., {Takagi} T., {Matsuhara} H., et~al., 2010{\natexlab{b}}, \aap, 514,
  A6

\bibitem[{Hwang} et~al.(2007){Hwang}, {Lee}, {Lee} et~al.]{2007ApJS..172..583H}
{Hwang} N., {Lee} M.~G., {Lee} H.~M., et~al., 2007, \apjs, 172, 583

\bibitem[{Jeon} et~al.(2010){Jeon}, {Im}, {Ibrahimov}, {Lee}, {Lee} \&
  {Lee}]{2010ApJS..190..166J}
{Jeon} Y., {Im} M., {Ibrahimov} M., {Lee} H.~M., {Lee} I., {Lee} M.~G., 2010,
  \apjs, 190, 166

\bibitem[{Karouzos} et~al.(2014){Karouzos}, {Im}, {Trichas}
  et~al.]{2014ApJ...784..137K}
{Karouzos} M., {Im} M., {Trichas} M., et~al., 2014, \apj, 784, 137

\bibitem[{Kilerci Eser} et~al.(2014){Kilerci Eser}, {Goto} \& {Doi}]{Ece2014}
{Kilerci Eser} E., {Goto} T., {Doi} Y., 2014, \mnras, 0, in press.

\bibitem[{Kim} et~al.(2012){Kim}, {Lee}, {Matsuhara}
  et~al.]{2012A&A...548A..29K}
{Kim} S.~J., {Lee} H.~M., {Matsuhara} H., et~al., 2012, \aap, 548, A29

\bibitem[{Koyama} et~al.(2008){Koyama}, {Kodama}, {Shimasaku}
  et~al.]{2008MNRAS.391.1758K}
{Koyama} Y., {Kodama} T., {Shimasaku} K., et~al., 2008, \mnras, 391, 1758

\bibitem[{Lagache} et~al.(1999){Lagache}, {Abergel}, {Boulanger}, {D{\'e}sert}
  \& {Puget}]{1999A&A...344..322L}
{Lagache} G., {Abergel} A., {Boulanger} F., {D{\'e}sert} F.~X., {Puget} J.-L.,
  1999, \aap, 344, 322

\bibitem[{Le Floc'h} et~al.(2005){Le Floc'h}, {Papovich}, {Dole}
  et~al.]{2005ApJ...632..169L}
{Le Floc'h} E., {Papovich} C., {Dole} H., et~al., 2005, \apj, 632, 169

\bibitem[{Lee} et~al.(2009){Lee}, {Kim}, {Im} et~al.]{2009PASJ...61..375L}
{Lee} H.~M., {Kim} S.~J., {Im} M., et~al., 2009, \pasj, 61, 375

\bibitem[{Miyazaki} et~al.(2012){Miyazaki}, {Komiyama}, {Nakaya}
  et~al.]{2012SPIE.8446E..0ZM}
{Miyazaki} S., {Komiyama} Y., {Nakaya} H., et~al., 2012, in { Society of
  Photo-Optical Instrumentation Engineers (SPIE) Conference Series\/}, vol.
  8446 of { Society of Photo-Optical Instrumentation Engineers (SPIE)
  Conference Series\/}

\bibitem[{Oi} et~al.(2014){Oi}, {Matsuhara}, {Murata} et~al.]{Oi2014}
{Oi} N., {Matsuhara} H., {Murata} K., et~al., 2014, \aap, 566, A60

\bibitem[{Shim} et~al.(2013){Shim}, {Im}, {Ko} et~al.]{2013ApJS..207...37S}
{Shim} H., {Im} M., {Ko} J., et~al., 2013, \apjs, 207, 37

\bibitem[{Takagi} et~al.(2010){Takagi}, {Ohyama}, {Goto} et~al.]{Takagi_PAH}
{Takagi} T., {Ohyama} Y., {Goto} T., et~al., 2010, \aap, 514, A5

\end{thebibliography}
\bibliographystyle{mnras}

\end{document}